\documentclass[a4paper,fleqn,usenatbib]{mnras}

\usepackage{newtxtext,newtxmath}

\usepackage[T1]{fontenc}
\usepackage{ae,aecompl}

\usepackage{graphicx}	
\usepackage{amsmath}	
\usepackage{amssymb}	

\title[Young Blue stragglers]{Young Blue Straggler Stars in the Galactic Field}

\author[Gemunu Ekanayake et al.]{
Gemunu Ekanayake,$^{1}$\thanks{E-mail: gemunu.ekanayake@mville.edu}
and Ronald Wilhelm$^{2}$
\\
$^{1}$Department of Physics, Manhattanville college, Purchase, NY 10577, USA\\
$^{2}$Department of Physics and Astronomy, University of Kentucky, Lexington, KY 40506, USA\\
}

\date{Accepted XXX. Received YYY; in original form ZZZ}

\pubyear{2018}

\begin{document}
\label{firstpage}
\pagerange{\pageref{firstpage}--\pageref{lastpage}}
\maketitle

\begin{abstract}
In this study we present an analysis of a sample of field blue straggler (BS) stars that show high ultra violet emission in their spectral energy distributions (SED): indication of a hot white dwarf (WD) companion to BS. Using photometry available in the  Sloan Digital Sky Survey (SDSS) and Galaxy Evolution Explorer (GALEX ) surveys we identified 80 stars with UV excess.
To determine the parameter distributions (mass, temperature and age) of the  WD companions, we developed a fitting routine
that could fit binary model SEDs to the observed SED. Results from this fit indicate the need for a hot WD companion to provide the excess UV flux. The WD mass distribution peaks at $\sim$$0.4\ M_{\odot}$, suggesting the primary formation channel of field BSs is case B mass transfer, i.e. when the donor star is in red giant phase of its evolution. Based on stellar evolutionary models, we estimate the lower limit of the binary mass transfer efficiency to be $\beta \sim 0.5$.
\end{abstract}




\section{Introduction}
Blue Straggler stars were first discovered in a photometric study of  globular cluster M3 by Sandage in 1953.\citep{bs5}.
They are identified by their position in color magnitude diagram, in which they appear
along the extension of the main sequence but bluer and brighter than the main sequence turn-off, giving the appearance of a younger stellar population.
The BS sequence is a typical feature of most globular clusters.
Since Sandage's discovery, blue stragglers have been observed
in many stellar environments, including open star clusters \citep{bs1}, globular clusters
\citep{bs2}, dwarf spheroidal galaxies \citep{bs3} and the Galactic field\citep{bs4}.

The exact mechanism for the formation of blue stragglers is still not fully understood. 
Single star evolutionary theory fails to explaining the existence of
 BS stars since it is not possible for stars to have masses greater than the main-sequence turnoff in a single
epoch star formation environment. Under the correct circumstances, however, it is possible for stars to gain significant mass long after the star formation epoch has ended.  Two leading theories rely on the basic idea that BSs are formed by adding mass to a main sequence star in a binary or  multiple stellar system via some interaction mechanism.
(1) Mass transfer between two stars in a binary system:
This is the case where the more massive star in a binary system transfers material, during its post
main-sequence phase, to its companion star.
Though still not fully understood McCrea \citep{bs6}  provided the first, seminal idea of mass transfer and its consequences during binary evolution.
If the mass transfer
is stable, this may add sufficient mass to the secondary to convert it into a blue
straggler. This appears to be the main formation channel for BS stars, particularly in the low density environment of the Galactic field. Convincing evidence for operation of this process was reported by  \cite{gosnell} for 3 stars in the old open cluster NGC 188.
(2) Merger between stellar systems:
 This could happen in various scenarios, such as the merger of a contact binary, a merger during a dynamical encounter and a
merger of an inner binary in hierarchical triple system. In such a scenario, it is expected that the merged stars will form a single more massive star.

Generally the mass transfer binaries are classified on the basis of the evolutionary phase of the donor \citep{bss2}.
\textbf{Case A:}
Mass transfer occurs during hydrogen core burning phase of the donor.This can happen if the orbital separation of the binary is small (usually an orbital
period of a few days).
\textbf{Case B:}
Mass transfer occurs after exhaustion of hydrogen in the core of the donor and the donor enters the red giant phase. In this case orbital period is about 100
days or less,
but significantly longer than case A.
\textbf{Case C:}
Mass transfer occurs after exhaustion of core helium (He) burning and the donor enters the AGB phase.
Here the orbital period is generally greater than 100 days.
It is believed that case A mass transfer is  more likely to be result in the coalescence of the two stars.\citep{bs7}
In that sense it can be treated as one of the merger scenarios.
According to the binary evolution simulations \citep{bs7}, in order to produce a BS via case B or case C, the mass transfer has to be
stable. During stable mass transfer the donor stays within its Roche-lobe. Therefore the stability of the mass transfer depends on donor's response to the mass loss, on how conservative the process is and on the angular momentum loss processes in this binary.
Given stable mass transfer, case B will result in a  He white dwarf companion bound to the blue straggler, while in 
case C the remnant companion is a CO white dwarf.
\subsection{Field Blue Straggler Stars}
For a given stellar population the dominant formation channel of BSs depends on its environment.
The BSs in the Galactic field, where the stellar number density is much lower than globular clusters,
are formed primarily from binary mass transfer \citep{bs4}. High resolution study of Field BSs
done by Preston an Sneden \citep{bs4} suggested that 60 \% their sample were binaries.
Moreover they concluded that the great majority of field BSs are probably created by
Roche-lobe overflow during red giant branch evolution (case B).
Ryan et al. \citep{bs17} studied lithium deficiency and rotation of the BSs and came to the conclusion
that these stars can be regarded as mass transfer binaries.The origin of BSs in the Galactic field is clearly tied to the overall formation of the Galactic halo.  Since star formation ended in the halo over 10 giga-years ago, most identified BSs originate through case B/C evolution.  The exception, as argued by \cite{bss3}, is a modest fraction of BSs that may actually be young, massive stars that have been accreted from Galactic satellites which contain young stellar populations.

To test the theory with observations, it is vital to identify the clean sample of mass transfer candidates.
Because blue stragglers in globular clusters are contaminated by those formed via collisions, field blue stragglers are the best candidates for
a clean sample of mass-transfer BSs.
As mentioned above the mass-transfer process results in a BS with a white dwarf companion.
BSs are much brighter than WDs at optical wavelengths so, such binaries are difficult to observe directly.
However, if the WD companions to BS are  sufficiently young and hot, they can be detected at ultraviolet wavelengths.
Studying a sample of such recent BS/WD systems can set constraints on the mass transfer formation mechanism. 
In this paper we present a study of field BSs that show UV excess in their SED, in order to characterize their mass transfer history. Sections 2 and 3 describe the methodology for identifying the UV excess stars by use of the SDSS and GALEX photometry. Section 4 summarizes our analysis and techniques used to determine the WD parameters.  A brief discussion and our conclusions are provided in section 5.

\section{Data}
Over the years many photometric studies have been done to identify field BSs, primarily using color-color plots. For example \cite{bss4} identified ~2700 field BSs using SDSS photometry. \cite{bss5} used SDSS photometry in combination with spectroscopy to identify field BSs.

The data used for this study were taken from the Sloan Digital Sky Survey Data Release 12 (SDSS DR12)
and Galaxy Evolution Explorer (GALEX) GR5, 	an ultraviolet survey.

\subsection{SDSS Data}
SDSS DR12 offers the latest data from SDSS project 3. It provides photometry in 5 bands (u, g, r, i and z) for
400 million objects and low resolution spectroscopy with resolution R $\sim$ 1800 and wavelength coverage 3800-9200A, for about 2 million objects.

Field BSs can be identified in a u-g vs g-r diagram by use of the following color cuts.
\begin{equation}
0.60 < (u-g) < 1.60
\end{equation}
\begin{equation}
-0.20 < (g-r) < 0.05
\end{equation}

This area in the color-color diagram is populated by A-type stars including high gravity BS and low gravity BHB
stars.
Therefore, in order to separate the BSs from BHB stars one would need stellar parameters for
individual stars. For that purpose we employed the current version of the Sloan Extension for Galactic Understanding
and Exploration (SEGUE) Stellar Parameter Pipeline (SSPP) \citep{bs10}.
We select only the BSs above 7000 K to avoid contamination of other F,G type main sequence stars and/or variable RR lyrae stars.

\subsection{GALEX Data}

GALEX photometry was performed in two ultraviolet(UV) bands Far-UV(FUV)
and Near-UV(NUV).
The effective wavelengths are 1516 and 2267 angstroms for FUV and NUV bands respectively.

\subsection{Cross-Identification of sources}

Most GALEX observations were designed to cover the SDSS footprint at a comparable depth. Constructing spectral energy distributions (SEDs) which range from
UV to IR requires combining SDSS sources to GALEX
counterparts.
The matching was done on line by use of the CDS X-Match Service \footnote{http://cdsxmatch.u-strasbg.fr/xmatch},
adopting a match radius of 5". Such a match highly depends on the
positional accuracy and resolution of both surveys. Because the GALEX images have lower angular resolution there are cases with multiple GALEX-matches per given SDSS source. 
These comprise about 10\% of our total
sample. In such cases we used only the closest distance matches.
In vast majority of cases  ($> 90\%$) GALEX and SDSS sources are matched within 2 arc-seconds.
Our final sample includes 2188 stars.

SDSS data were corrected for interstellar extinction using the extinction relations given in \cite{bs15} while FUV and NUV were corrected for extinction using the relations given in \cite{bs14}.

\section{UV-Excess Stars}
Identification of UV-excess stars based on their position in FUV/optical color-color plot is one useful application of UV photometry.
Using GALEX and SDSS colors together have enabled the discovery of white dwarf-main sequence
(WDMS) binary systems, i.e., binaries with WD primaries and late-type main-sequence secondaries \citep{bss6}

 \cite{bs12} adopted a similar approach to identify the FGK-type stars with excess UV in their spectral energy distribution.
They used combined data from GALEX satellite's
far-UV (FUV) and near-UV (NUV) bandpasses as well as from the ground-based SDSS survey and
the Kepler Input Catalog to identify  stars that exhibit FUV-excesses relative to their NUV fluxes and spectral types.
They considered  that these UV excesses originate from various types of hot stars, including
white dwarf DA and sdB stars, binaries, and strong chromosphere stars that are young or in active
binaries.They calibrate the UV-excess stars using their distribution in (FUV - NUV) - $T_{eff}$ plane (Figure 3 \cite{bs12}).
We adopted a similar method in selecting BS UV-excess stars.
\begin{figure}
\includegraphics[width=\columnwidth]{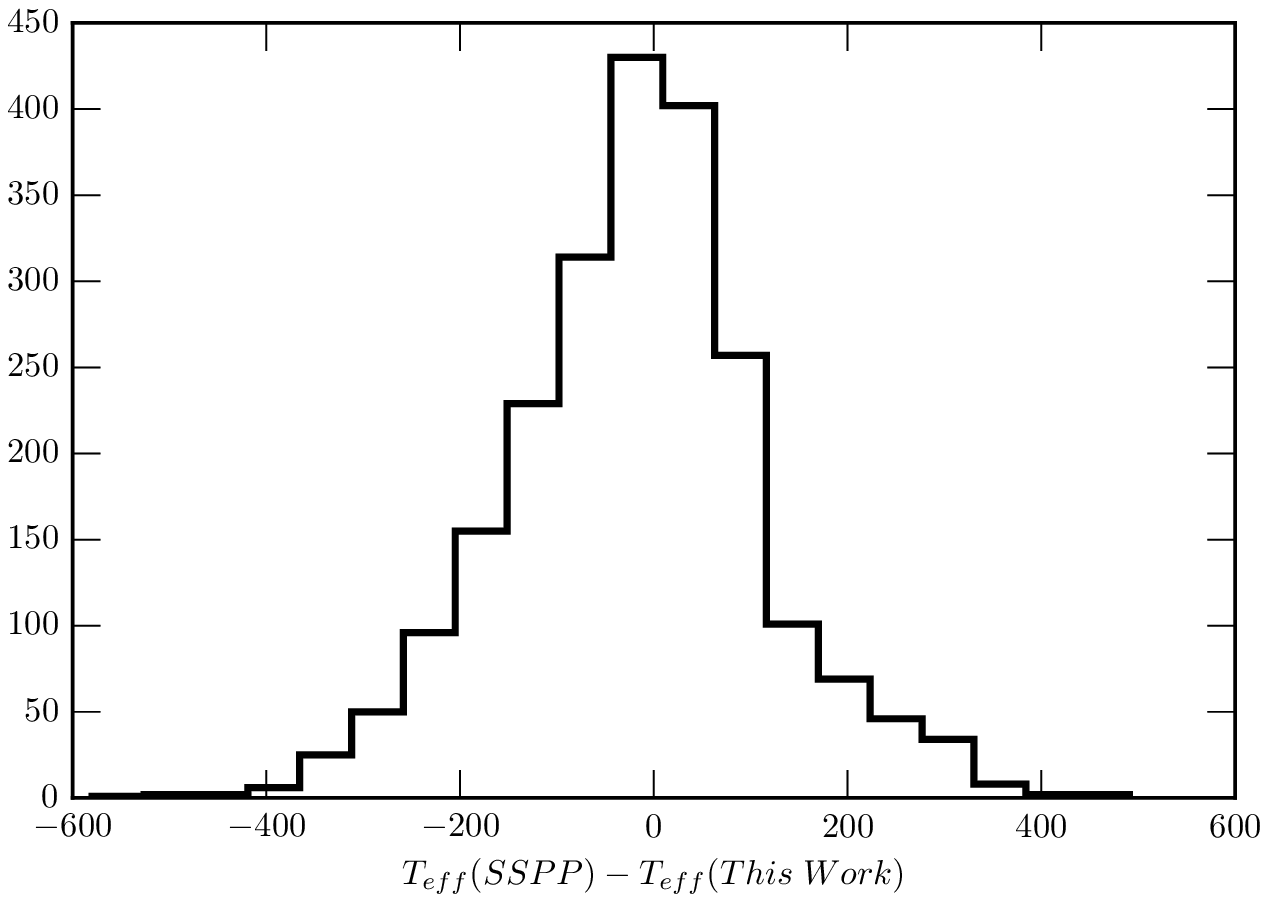}
\includegraphics[width=\columnwidth]{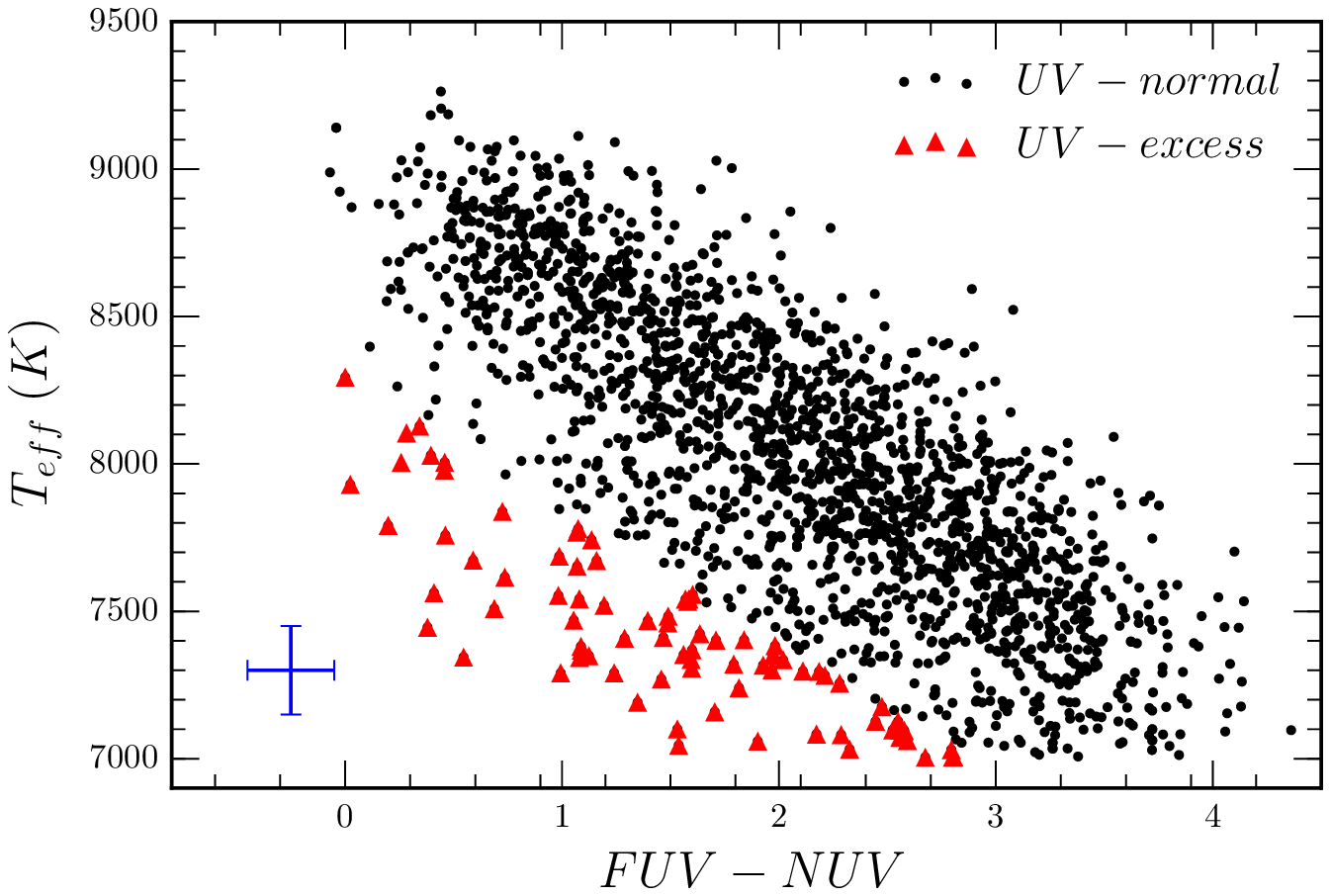}
\caption{Top: Comparison of temperatures of BSs in this work with SSPP. Bottom: s(FUV-NUV) vs. SSPP temperature values from SDSS-GALEX data. Data represented by  red triangles deviate more than $2\sigma$  from the fit to the main diagonal sequence and identified as BSs with UV-excess.}
\end{figure}

\subsection{Temperature Sanity Check}
The method for selecting UV excess BSs depend on the temperatures of the stars.
Rather than simply adopting the SSPP $T_{eff}$ values directly,
we remeasured the temperatures for our sample stars by fitting of the Hydrogen Balmer lines of each SDSS spectrum to model synthetic spectra.
A grid of synthetic spectra, spanning the stellar parameter range of our sample, was generated using Kurucz ATLAS12 atmosphere models \footnote{Kurucz models
are available at http:kurucz.harvard.edu} and the spectral synthesis routine, SPECTRUM \citep{spec}.  In order to achieve a finer grid resolution we
interpolate the Kurucz models
using kmod IDL package.\footnote{http://www.as.utexas.edu/~hebe/}
kmod interpolates linearly a Kurucz model for the desired values of effective temperature, surface gravity and metallicity using 8 surrounding models.
Our final grid consists of spectra with the temperatures in the range 6500 - 10,000 K in steps of 125 K.
Finally the point estimates of mean and error in temperatures were calculated. The mean error  obtained for temperatures is 140 K. We compared the temperatures obtained from our method to those obtained from
the SSPP in order to check the consistency (Figure 1(a)). Both SSPP and our method
predicts consistent values for the majority of the stars. We excluded from our sample any 
stars which have a $T_{eff}$ difference larger than 250 K.  Given the larger uncertainties in our
measurements, we have adopted the SSPP $T_{eff}$ values for our final sample. 

We plotted $T_{eff}$ values determined from the SSPP vs (FUV - NUV) color of the stars in our sample (Figure 1). The diagonal sequence in the Figure 1 reflects the relationship between SDSS $T_{eff}$ and (FUV - NUV) for the main-sequence stars. We fit the main diagonal sequence with a quadratic fit and we identify stars lying more than 2 sigma below the regression as UV-excess BS stars. These are our candidates for the BSs with a hot WD companion.  Their GALEX and SDSS photometry along with their stellar parameters are given in Table A1.

\section{BS-WD  SED Fitting}

In order to obtain the stellar parameters of BS-WD binaries, we fit the observed SEDs of UV-excess stars with
a composite model SEDs composed of BSs and white dwarfs. We employed a theoretical grid of spectra 
\citep{kuru04} with $ 6000  < T_{eff} < 10000 $ K 
in steps of 250 K, $0.5 < log g< 5.0 $ in steps
of 0.5, and $-2.5 < [Fe/H] < 0.5$ in steps of 0.5.

For white dwarf models we employed the theoretical color tables developed by Bergeron( University of Montreal; private communication).
For these models stellar masses and cooling ages are obtained from a detailed evolutionary cooling sequences appropriate for these stars. For the white dwarfs with pure hydrogen model atmospheres above temperatures 30,000 K, the carbon-core cooling models of Wood \citep{paka30} with thick hydrogen layers of $M_{H}/M* = 10^{-4}$ were used.
 For temperatures below 30,000 K, cooling models similar to those of  \cite{paka31}
 but with carbon-oxygen cores and $M_{H}/M* = 10^{-4}$ were used.
 
 \subsection{Synthetic Fluxes}

In the course of fitting procedure we calculate synthetic fluxes for both BS and WD in GALEX and SDSSS bandpasses.
The synthetic spectra can be converted to a monochromatic flux  for a given bandpass via,

\begin{equation}
  f = \frac{\int_{\lambda_{i}}^{\lambda_{f}} \lambda f_{\lambda} S_{\lambda} d \lambda}
  {\int_{\lambda_{i}}^{\lambda_{f}} \lambda S_{\lambda} d \lambda}
\end{equation}
\

where $f_{\lambda}$ is the flux at a given wavelength $\lambda $  and
$S_{\lambda} $ is the filter response at a given wavelength.
The integration limits are the minimum and maximum wavelength of the bandpass.

\subsection{Observed Fluxes}

The observed, extinction corrected, magnitudes  were transformed to fluxes using the standard formulas.
For SDSS bandpasses \citep{bss7},
\begin{equation}
  m_{AB} = -2.5\log f_{\nu} - 48.6
\end{equation}

The FUV and NUV fluxes are determined by means of the conversion,

\begin{equation}
  m_{FUV} = -2.5\log \frac{Flux_{FUV}}{1.40 * 10^{-15}} + 18.82
\end{equation}

\begin{equation}
  m_{NUV} = -2.5\log \frac{Flux_{NUV}}{2.06 * 10^{-16}} + 20.08
\end{equation}

\subsection{SED Fitting}

\begin{figure*}
\includegraphics[width=\textwidth]{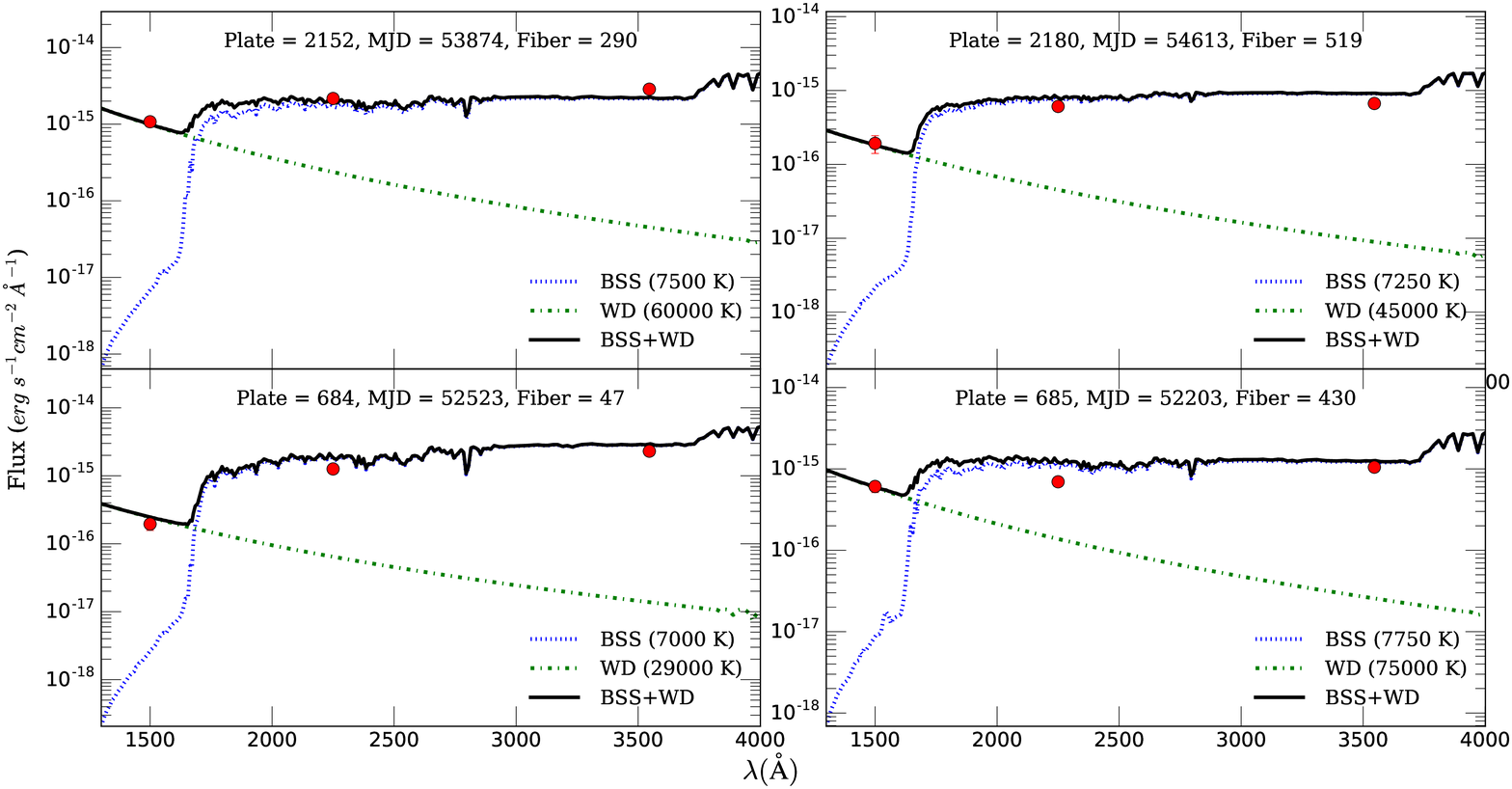}
\includegraphics[width=\textwidth]{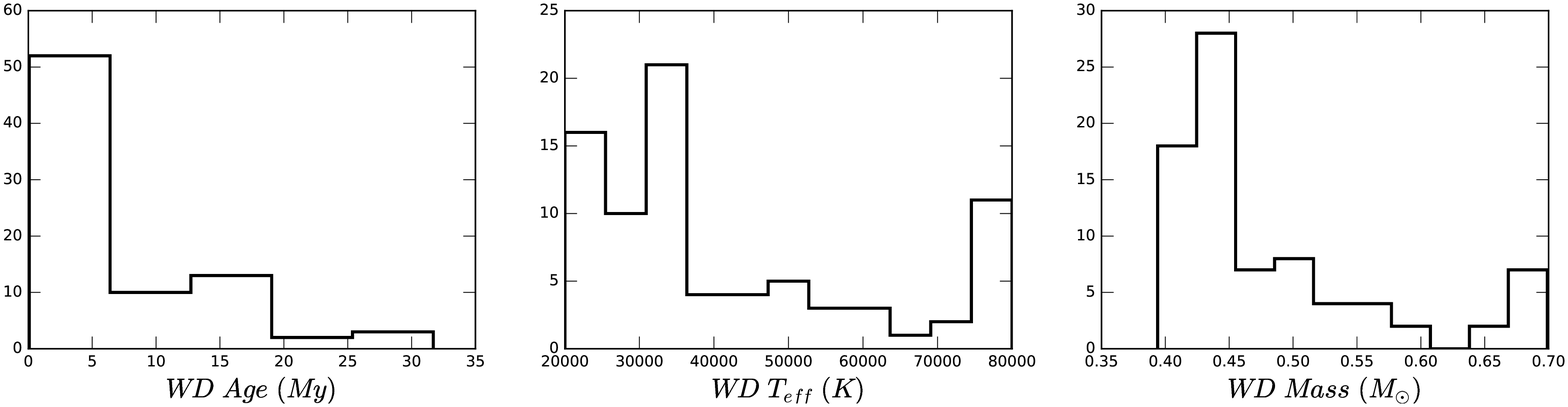}
\caption{Top panel: Examples of BS+WD fitting. Red circles are the observed FUV,NUV and u band fluxes. Blue and green dashed lines represent the synthetic spectra of BS and WD respectively. Black solid line is the composite (BS+WD)  best fit spectrum to the observed fluxes. Bottom panel: White Dwarf parameter distributions obtained from SED fitting.}
\end{figure*}
In the fitting process we minimize the following chi-square function:

\begin{equation}
\chi^2=\sum\limits_{i}\frac{(\alpha^2f_{wd,i}+\beta^2f_{bs,i}-F_{i})^2}{\sigma_{F,i}^2}
\end{equation}
\\
Where $i$ sum over all bandpasses.

$f_{bs,i},f_{wd,i},F_{i}$ are the flux of the BS model,
flux of the WD model and flux of the observed star respectively.
 $\alpha= (R_{bs}/D)$ and $\beta= (R_{wd}/D)$ are scale factors depend on radii($R_{bs}\ and\ R_{wd}$) of each component and the distance ($D$).

The radii of both components ($R_{bs},R_{wd}$) and distance are necessary as scale factors for the individual
fluxes when combining the atmosphere models of both components to a single
SED.
However, for WD models the mass and the surface gravity is known. So for each model in the grid we can calculate
the radius, according to:

\begin{equation}
  R_{WD} = \sqrt{\frac{G M_{WD}}{g_{WD}}}
\end{equation}
\\
For the BS, we first obtain absolute magnitude calibration for our sample.
we used transformation equations given by \cite{bss8}, which were derived from the SDSS
stars with known UBVRI photometry, including a sample of BSS stars.

\begin{equation}
  V = g - 0.561 (g-r) - 0.004
\end{equation}

\begin{equation}
    (B-V) = 0.916 (g-r) + 0.187
\end{equation}
\

These transformations are valid within the range $-0.5 < g-r <1.0$, \citep{bss10}
which is consistent with the color range we adopted here.

Using the V magnitude we obtained from equation 9, we can now calculate the absolute magnitude in V band ($M_{v}$) using the
relation given by \cite{bss9}:
\begin{equation}
  M_{v} = 1.32 + 4.05 (B-V) - 0.45 \left[\frac{Fe}{H}\right]
\end{equation}

This relation was constructed by studying the BSs in  globular clusters with different metallicities covering a wide range of colors and absolute magnitudes.

The luminosity of the star can be found by use of $M_{v}$ and bolometric correction in equation 12.

\begin{equation}
  \left(\frac{L}{L_{o}}\right) = 10 ^{-0.4 \left[M_{v} - V_{o} - 31.752 + (BC_{v} - BC_{v,o})\right]}
\end{equation}
\

where we adopt the bolometric corrections  given by \cite{bs13}.

Then, the radius of the BS can be estimated by use of the relation,

\begin{equation}
  \left(\frac{R}{R_{o}}\right) = \left(\frac{T_{o}}{T_{eff}}\right)^2 \left(\frac{L}{L_{o}}\right)^{0.5}
\end{equation}

To select the best fit white dwarf model, the $\chi^2$ value of each fit is calculated by use of equation 7.
The optimal SED fits for four stars are shown in the top panel of Figure 2.
The presence of a hot white dwarf companion in a BS binary causes the excess FUV emission of the system. This is clear in Figure 2. The expected emission from BSs without WD companions is much fainter than the observed flux  at FUV. Adding WD companions of increasing temperature results in bright, blue emission as evidence by the best fitting composite model (Black line in Figure 2: Top panel).

\subsection{WD Parameters}
 The distributions of mass, age and effective temperatures of the white dwarfs determined from our SED fitting routine are shown in the bottom panel of Figure 2.
The most striking features of the distributions are:
Vast majority of white dwarfs are very young (few million years old) and  with temperatures range between 20000 - 40000 K (See bottom panel of figure 2).
This is consistent with our initial selection criteria of BS, as recently formed BS are expected to have much larger UV excess.
One interesting feature of the mass distribution is the peak at white dwarf mass  ~0.43 $M_{\odot}$.
White dwarf stars with masses  below 0.47 $M_{\odot}$ are thought to be He-core white dwarfs. ( Core helium ignition starts when the core mass is roughly
0.47$M_{\odot}$)
At the stage when mass transfer occurs, the expanding red giant core has not yet fully grown, therefore the resultant white dwarf will have a lower mass.
If these WDs  were single stars, they would have main sequence life times greater than age of the Universe.
Therefore, these systems must be products of case B mass transfer, the only possible solution for 
producing such low mass white dwarfs.

On the other hand white dwarfs with masses $ > 0.5 M_{\odot}$ are consistent with predictions from the case C, i.e., mass transfer from an asymptotic giant to a main sequence star which produces a carbon - oxygen white dwarf companion with mass of $\sim$ .5 $M_{\odot}$ to 0.6 $M_{\odot}$ dictated by the core mass of the asymptotic giant donor  at the end of the mass transfer phase \citep{bss1}.
The WD Mass distribution in the Galactic field peaks at about 0.59  $M_{\odot}$ and exhibits a significant low- mass tail of white dwarfs with masses lower than 0.45 $M_{\odot}$ which peaks at 0.40 $M_{\odot}$.
These white dwarfs are predominantly found in close binary systems, mostly with another white dwarf or a neutron star companion \citep{paka2}.
Interestingly, the low mass end of this distribution is consistent with our results as well.

\subsection{Mass Transfer efficiency}

Mass transfer efficiency $\beta$, is defined as the mass fraction of the lost mass from the primary accreted by the secondary. 
This is an important parameter for binary evolution calculations where $\beta$ is frequently treated with a constant value due to its large uncertainty.\citep{mt1, bs7}.
 \cite{lu2010} used Monte Carlo simulations to investigate the origin of BS population in M67. In their calculations they used the $\beta = 0.5$ and $\beta = 1.0$ (fully conservative mass transfer, i.e. no mass or angular momentum loss from the system) and found that the higher value could reproduce the data better.
Our WD parameters obtained via the BS-WD fitting can be used to infer a lower limit for $\beta$ in our current sample.

\begin{figure}
\includegraphics[width=\columnwidth]{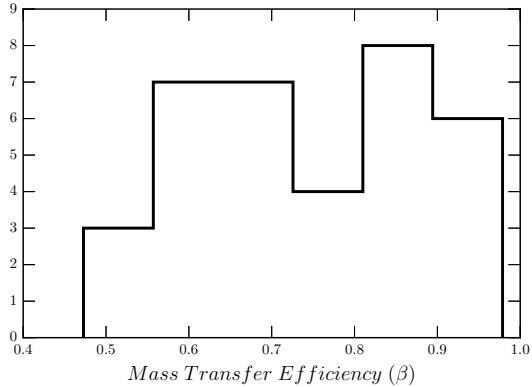} 
\caption{Histogram of the lower limits of the mass transfer efficiencies. }
\end{figure}

The amount of mass transferred from the now WD to BS, $\delta M_{t}$,
\begin{equation}
 	\delta M_{trans} =  M_{i} - M_{wd}
\end{equation}

where $M_{i}$ is the progenitor mass of the WD.

We interpolate theoretical relations which were constructed using BaSTI evolutionary models \citep{paka32} so that we can calculate the progenitor mass at a given metallicity and age .
We adopted the non-canonical BaSTI models with the mass loss efficiency of the Reimers law(Reimers 1977) set to $\eta$ =0.2.
The initial He mass fraction ranges from 0.245 to 0.303, for the more metal-poor to the
more metal-rich composition, respectively. 

\cite{paka26} estimated the  ages of  field halo stars using a sample of
SDSS stars. They estimated a mean age 11 Gy, which we adopted in our calculations to determine the
mass from interpolated BASTI models.
The field stars of \cite{paka26} have a mean metallicity of -1.7 dex. To be consistent with that work
we limited our sample to the stars with $[Fe/H]$ $<$ -1.35. This reduces our sample to 35 stars.

To calculate the amount of mass transferred, the initial mass of the secondary star (Now the BS) is required.  This initial mass value cannot be readily
inferred, without first knowing the value of $\beta$.
But we can estimate the lower limit of the transferred mass by considering the halo turnoff mass of 0.8 $M_{\odot}$,
Then the mass accreted on to the secondary (now BS) is given by,

\begin{equation}
 	\delta M_{acc} =  M_{BS} - 0.8
\end{equation}

Ratio of equations 15 and 14  will give the lower limit of the mass transfer efficiency.
The derived $\beta$  values are shown as a histogram  in Figure 3. All the stars have mass transfer efficiencies
 above  $\beta \sim 0.5$.
 The value we obtained for the lower limit of $\beta$ provides clues about the nature of these ancient, low-mass binary systems. 
The tendency for our systems to favor a more conservative mass tranfer (larger $\beta$), should help to inform tranfer models 
where donor initial mass less than 1.0 $M_{\odot}$.    

\section{Summary}
We utilized UV-optical SED study of the BS binaries in the Galactic field.
Using our fitting routine we identified WD companions of field BSs formed by mass transfer.
We found that these BSs have significant FUV excess by comparing the observed SED to a composite model with WD and BS components.

We found that our sample WDs range in age from a few million to a few tens of millions of years old, suggesting mass transfer in these binaries ended
relatively recently. In addition   we conclude that majority of the WD companions are helium WDs. Mass transfer from RGB stars in binary systems is the obvious way to produce such low mass stars.

Combining the WD stellar parameters with the evolutionary models we estimated the lower limit of
the binary mass transfer efficiency in these stars to be, $\beta = 0.5$.

In order to determine the exact formation mechanisms for a particular population of BSs a detailed
characterization of the BSs is needed. This requires determining the rotation velocities and binary
orbital parameters.
Since the formation channel for BS binaries can distinguish from the type of companion star expected,
it is vital to identify the observational constraints of the companion. These binaries will be good test cases for binary mass transfer modeling efforts
in the future.

\section*{Acknowledgements}
The authors would like to thank our referee, George Preston, for his insightful comments and suggestions which 
improved the overall content and flow of this paper.\\
This work made use of the public data services provided
by MAST and SDSS.

\newpage

\appendix

\section{}
%

%
%


\begin{table}
\centering
\caption{Stellar parameters and photometry of UV-excess BSs}
\resizebox{\textwidth}{!}{%
\begin{tabular}{@{}lllllllllll@{}}
\hline
SDSS ID             & $T_{eff}$ & $log\ g$    &   $[Fe/H]$  & FUV          & NUV          & u            & g            & r            & i            & z            \\ \hline
915-52443-549  & 7423 $\pm$\ 157 & 3.84 $\pm$\ 0.49 & -0.69 $\pm$\ 0.2  & 19.0 $\pm$\ 0.17  & 17.13 $\pm$\ 0.04 & 15.76 $\pm$\ 0.02 & 14.71 $\pm$\ 0.01 & 14.68 $\pm$\ 0.02 & 14.73 $\pm$\ 0.01 & 14.82 $\pm$\ 0.02 \\
740-52263-440  & 7673 $\pm$\ 129 & 4.32 $\pm$\ 0.15 & -1.98 $\pm$\ 0.07 & 19.89 $\pm$\ 0.24 & 19.35 $\pm$\ 0.13 & 18.64 $\pm$\ 0.02 & 17.84 $\pm$\ 0.01 & 17.89 $\pm$\ 0.01 & 17.94 $\pm$\ 0.02 & 18.07 $\pm$\ 0.03 \\
684-52523-47   & 7004 $\pm$\ 40  & 3.88 $\pm$\ 0.23 & -0.78 $\pm$\ 0.05 & 20.93 $\pm$\ 0.21 & 18.06 $\pm$\ 0.03 & 16.41 $\pm$\ 0.03 & 15.44 $\pm$\ 0.02 & 15.3 $\pm$\ 0.01  & 15.27 $\pm$\ 0.02 & 15.34 $\pm$\ 0.02 \\
3131-54731-429 & 7936 $\pm$\ 110 & 4.36 $\pm$\ 0.03 & -1.35 $\pm$\ 0.1  & 20.08 $\pm$\ 0.25 & 19.57 $\pm$\ 0.07 & 18.55 $\pm$\ 0.03 & 17.53 $\pm$\ 0.02 & 17.56 $\pm$\ 0.01 & 17.62 $\pm$\ 0.01 & 17.59 $\pm$\ 0.03 \\
2951-54592-114 & 7615 $\pm$\ 73  & 3.93 $\pm$\ 0.44 & -2.07 $\pm$\ 0.16 & 19.42 $\pm$\ 0.24 & 18.64 $\pm$\ 0.1  & 17.55 $\pm$\ 0.02 & 16.51 $\pm$\ 0.02 & 16.49 $\pm$\ 0.01 & 16.54 $\pm$\ 0.01 & 16.57 $\pm$\ 0.02 \\
2849-54454-349 & 7083 $\pm$\ 37  & 4.21 $\pm$\ 0.29 & -1.53 $\pm$\ 0.02 & 19.46 $\pm$\ 0.18 & 17.57 $\pm$\ 0.05 & 16.22 $\pm$\ 0.01 & 15.32 $\pm$\ 0.02 & 15.2 $\pm$\ 0.02  & 15.16 $\pm$\ 0.01 & 15.22 $\pm$\ 0.02 \\
1894-53240-211 & 7441 $\pm$\ 78  & 4.19 $\pm$\ 0.21 & -1.25 $\pm$\ 0.08 & 21.18 $\pm$\ 0.26 & 19.58 $\pm$\ 0.04 & 18.48 $\pm$\ 0.02 & 17.44 $\pm$\ 0.02 & 17.33 $\pm$\ 0.01 & 17.37 $\pm$\ 0.01 & 17.43 $\pm$\ 0.02 \\
2299-53711-626 & 7742 $\pm$\ 64  & 3.93 $\pm$\ 0.18 & -0.25 $\pm$\ 0.04 & 21.67 $\pm$\ 0.43 & 20.53 $\pm$\ 0.19 & 18.71 $\pm$\ 0.03 & 17.61 $\pm$\ 0.01 & 17.61 $\pm$\ 0.01 & 17.7 $\pm$\ 0.02  & 17.81 $\pm$\ 0.03 \\
2299-53711-453 & 7460 $\pm$\ 50  & 3.88 $\pm$\ 0.21 & -0.24 $\pm$\ 0.08 & 21.01 $\pm$\ 0.48 & 19.53 $\pm$\ 0.16 & 17.69 $\pm$\ 0.02 & 16.5 $\pm$\ 0.01  & 16.45 $\pm$\ 0.01 & 16.5 $\pm$\ 0.01  & 16.58 $\pm$\ 0.01 \\
2848-54453-473 & 7032 $\pm$\ 31  & 4.14 $\pm$\ 0.25 & -1.41 $\pm$\ 0.04 & 19.47 $\pm$\ 0.1  & 17.15 $\pm$\ 0.02 & 15.9 $\pm$\ 0.01  & 14.95 $\pm$\ 0.02 & 14.82 $\pm$\ 0.02 & 14.81 $\pm$\ 0.02 & 14.88 $\pm$\ 0.01 \\
1254-52972-515 & 7672 $\pm$\ 140 & 4.2 $\pm$\ 0.11  & -0.02 $\pm$\ 0.05 & 18.66 $\pm$\ 0.38 & 17.5 $\pm$\ 0.12  & 15.96 $\pm$\ 0.02 & 15.01 $\pm$\ 0.01 & 15.07 $\pm$\ 0.02 & 15.15 $\pm$\ 0.01 & 15.33 $\pm$\ 0.01 \\
1252-52970-306 & 7791 $\pm$\ 0   & 4.14 $\pm$\ 0.23 & -0.25 $\pm$\ 0.03 & 16.95 $\pm$\ 0.11 & 16.75 $\pm$\ 0.06 & 16.28 $\pm$\ 0.02 & 15.4 $\pm$\ 0.02  & 15.41 $\pm$\ 0.01 & 15.48 $\pm$\ 0.01 & 15.61 $\pm$\ 0.02 \\
2335-53730-480 & 7468 $\pm$\ 33  & 3.98 $\pm$\ 0.31 & -0.49 $\pm$\ 0.14 & 19.14 $\pm$\ 0.46 & 17.75 $\pm$\ 0.15 & 16.04 $\pm$\ 0.05 & 14.98 $\pm$\ 0.02 & 14.9 $\pm$\ 0.02  & 14.97 $\pm$\ 0.02 & 15.02 $\pm$\ 0.02 \\
3241-54884-335 & 8029 $\pm$\ 71  & 4.13 $\pm$\ 0.2  & -1.97 $\pm$\ 0.25 & 19.3 $\pm$\ 0.21  & 18.9 $\pm$\ 0.13  & 18.16 $\pm$\ 0.04 & 17.23 $\pm$\ 0.02 & 17.32 $\pm$\ 0.01 & 17.42 $\pm$\ 0.01 & 17.54 $\pm$\ 0.02 \\
685-52203-430  & 7686 $\pm$\ 112 & 3.93 $\pm$\ 0.36 & -1.03 $\pm$\ 0.07 & 19.69 $\pm$\ 0.17 & 18.7 $\pm$\ 0.08  & 17.26 $\pm$\ 0.01 & 16.11 $\pm$\ 0.02 & 16.09 $\pm$\ 0.02 & 16.12 $\pm$\ 0.02 & 16.15 $\pm$\ 0.02 \\
3130-54740-107 & 7344 $\pm$\ 69  & 4.15 $\pm$\ 0.22 & -1.89 $\pm$\ 0.07 & 19.87 $\pm$\ 0.19 & 19.32 $\pm$\ 0.07 & 18.45 $\pm$\ 0.02 & 17.45 $\pm$\ 0.02 & 17.38 $\pm$\ 0.01 & 17.35 $\pm$\ 0.02 & 17.43 $\pm$\ 0.02 \\
2252-53565-281 & 7301 $\pm$\ 44  & 4.19 $\pm$\ 0.16 & -0.27 $\pm$\ 0.04 & 21.02 $\pm$\ 0.39 & 19.05 $\pm$\ 0.05 & 17.4 $\pm$\ 0.01  & 16.32 $\pm$\ 0.02 & 16.24 $\pm$\ 0.01 & 16.26 $\pm$\ 0.01 & 16.3 $\pm$\ 0.01  \\
1961-53299-168 & 7291 $\pm$\ 45  & 4.05 $\pm$\ 0.19 & -0.76 $\pm$\ 0.02 & 19.49 $\pm$\ 0.17 & 18.25 $\pm$\ 0.07 & 17.61 $\pm$\ 0.02 & 16.61 $\pm$\ 0.01 & 16.53 $\pm$\ 0.02 & 16.54 $\pm$\ 0.02 & 16.62 $\pm$\ 0.02 \\
1960-53289-343 & 7127 $\pm$\ 33  & 4.23 $\pm$\ 0.17 & -0.54 $\pm$\ 0.06 & 20.9 $\pm$\ 0.4   & 18.45 $\pm$\ 0.09 & 16.71 $\pm$\ 0.02 & 15.68 $\pm$\ 0.02 & 15.54 $\pm$\ 0.01 & 15.54 $\pm$\ 0.02 & 15.57 $\pm$\ 0.02 \\
3138-54740-521 & 7381 $\pm$\ 31  & 4.03 $\pm$\ 0.3  & -1.11 $\pm$\ 0.03 & 21.04 $\pm$\ 0.45 & 19.95 $\pm$\ 0.19 & 18.44 $\pm$\ 0.03 & 17.4 $\pm$\ 0.01  & 17.31 $\pm$\ 0.01 & 17.35 $\pm$\ 0.02 & 17.38 $\pm$\ 0.02 \\
3138-54740-401 & 7562 $\pm$\ 177 & 4.31 $\pm$\ 0.16 & -2.59 $\pm$\ 0.01 & 19.49 $\pm$\ 0.21 & 19.08 $\pm$\ 0.12 & 17.77 $\pm$\ 0.03 & 16.76 $\pm$\ 0.01 & 16.75 $\pm$\ 0.01 & 16.73 $\pm$\ 0.01 & 16.71 $\pm$\ 0.02 \\
1857-53182-27  & 7343 $\pm$\ 129 & 4.31 $\pm$\ 0.2  & -0.88 $\pm$\ 0.08 & 21.53 $\pm$\ 0.34 & 20.45 $\pm$\ 0.19 & 18.87 $\pm$\ 0.03 & 17.9 $\pm$\ 0.01  & 17.77 $\pm$\ 0.01 & 17.79 $\pm$\ 0.01 & 17.88 $\pm$\ 0.03 \\
366-52017-28   & 7320 $\pm$\ 60  & 4.2 $\pm$\ 0.23  & -1.01 $\pm$\ 0.08 & 21.16 $\pm$\ 0.37 & 19.37 $\pm$\ 0.12 & 17.88 $\pm$\ 0.02 & 16.91 $\pm$\ 0.02 & 16.86 $\pm$\ 0.01 & 16.86 $\pm$\ 0.02 & 16.92 $\pm$\ 0.02 \\
2797-54616-614 & 8005 $\pm$\ 47  & 4.24 $\pm$\ 0.19 & -1.22 $\pm$\ 0.32 & 20.87 $\pm$\ 0.38 & 20.41 $\pm$\ 0.26 & 18.74 $\pm$\ 0.04 & 17.67 $\pm$\ 0.01 & 17.71 $\pm$\ 0.01 & 17.81 $\pm$\ 0.02 & 17.92 $\pm$\ 0.03 \\
2551-54552-257 & 7368 $\pm$\ 100 & 4.07 $\pm$\ 0.06 & -1.31 $\pm$\ 0.06 & 21.51 $\pm$\ 0.34 & 19.91 $\pm$\ 0.12 & 18.75 $\pm$\ 0.02 & 17.58 $\pm$\ 0.02 & 17.51 $\pm$\ 0.02 & 17.52 $\pm$\ 0.02 & 17.52 $\pm$\ 0.03 \\
2247-53857-96  & 7290 $\pm$\ 68  & 4.18 $\pm$\ 0.29 & -0.54 $\pm$\ 0.04 & 20.57 $\pm$\ 0.34 & 19.57 $\pm$\ 0.14 & 18.18 $\pm$\ 0.02 & 17.1 $\pm$\ 0.02  & 17.05 $\pm$\ 0.01 & 17.08 $\pm$\ 0.01 & 17.15 $\pm$\ 0.02 \\
2180-54613-519 & 7368 $\pm$\ 46  & 3.98 $\pm$\ 0.31 & -1.83 $\pm$\ 0.01 & 20.94 $\pm$\ 0.29 & 18.85 $\pm$\ 0.06 & 17.75 $\pm$\ 0.02 & 16.89 $\pm$\ 0.02 & 16.87 $\pm$\ 0.02 & 16.87 $\pm$\ 0.02 & 16.92 $\pm$\ 0.02 \\
1659-53224-452 & 7530 $\pm$\ 40  & 4.05 $\pm$\ 0.19 & -0.56 $\pm$\ 0.09 & 19.56 $\pm$\ 0.15 & 17.9 $\pm$\ 0.05  & 16.46 $\pm$\ 0.02 & 15.39 $\pm$\ 0.01 & 15.38 $\pm$\ 0.01 & 15.45 $\pm$\ 0.02 & 15.55 $\pm$\ 0.02 \\
2550-54206-232 & 7097 $\pm$\ 80  & 4.24 $\pm$\ 0.14 & -0.97 $\pm$\ 0.05 & 21.17 $\pm$\ 0.36 & 18.61 $\pm$\ 0.07 & 17.16 $\pm$\ 0.02 & 16.15 $\pm$\ 0.01 & 16.04 $\pm$\ 0.02 & 16.02 $\pm$\ 0.02 & 16.04 $\pm$\ 0.02 \\
2189-54624-640 & 7307 $\pm$\ 61  & 4.0 $\pm$\ 0.67  & -1.5 $\pm$\ 0.0   & 21.51 $\pm$\ 0.45 & 19.91 $\pm$\ 0.11 & 18.71 $\pm$\ 0.03 & 17.74 $\pm$\ 0.02 & 17.73 $\pm$\ 0.01 & 17.75 $\pm$\ 0.01 & 17.84 $\pm$\ 0.03 \\
595-52023-92   & 7542 $\pm$\ 81  & 4.11 $\pm$\ 0.08 & -1.72 $\pm$\ 0.04 & 19.07 $\pm$\ 0.18 & 17.99 $\pm$\ 0.05 & 16.88 $\pm$\ 0.02 & 15.89 $\pm$\ 0.01 & 15.87 $\pm$\ 0.01 & 15.92 $\pm$\ 0.01 & 15.98 $\pm$\ 0.02 \\
1727-53859-288 & 7045 $\pm$\ 39  & 3.84 $\pm$\ 0.43 & -0.53 $\pm$\ 0.02 & 21.79 $\pm$\ 0.43 & 20.25 $\pm$\ 0.14 & 18.61 $\pm$\ 0.02 & 17.52 $\pm$\ 0.02 & 17.42 $\pm$\ 0.02 & 17.42 $\pm$\ 0.02 & 17.48 $\pm$\ 0.02 \\
2782-54592-113 & 7445 $\pm$\ 103 & 3.88 $\pm$\ 0.45 & -1.61 $\pm$\ 0.05 & 19.64 $\pm$\ 0.21 & 19.26 $\pm$\ 0.12 & 18.33 $\pm$\ 0.02 & 17.23 $\pm$\ 0.01 & 17.16 $\pm$\ 0.01 & 17.12 $\pm$\ 0.01 & 17.11 $\pm$\ 0.02 \\
3308-54919-447 & 7368 $\pm$\ 82  & 3.92 $\pm$\ 0.4  & -1.41 $\pm$\ 0.09 & 20.9 $\pm$\ 0.37  & 19.81 $\pm$\ 0.14 & 18.8 $\pm$\ 0.03  & 17.95 $\pm$\ 0.02 & 17.91 $\pm$\ 0.01 & 17.96 $\pm$\ 0.02 & 18.04 $\pm$\ 0.03 \\
2781-54266-417 & 7768 $\pm$\ 145 & 4.37 $\pm$\ 0.16 & -1.34 $\pm$\ 0.03 & 20.35 $\pm$\ 0.26 & 19.28 $\pm$\ 0.11 & 18.16 $\pm$\ 0.02 & 17.23 $\pm$\ 0.02 & 17.31 $\pm$\ 0.01 & 17.34 $\pm$\ 0.01 & 17.41 $\pm$\ 0.02 \\
2156-54525-520 & 7483 $\pm$\ 58  & 3.89 $\pm$\ 0.42 & -0.98 $\pm$\ 0.02 & 21.75 $\pm$\ 0.42 & 20.26 $\pm$\ 0.14 & 18.71 $\pm$\ 0.02 & 17.7 $\pm$\ 0.01  & 17.68 $\pm$\ 0.02 & 17.74 $\pm$\ 0.01 & 17.83 $\pm$\ 0.02 \\
3297-54941-411 & 7411 $\pm$\ 56  & 3.98 $\pm$\ 0.45 & -1.14 $\pm$\ 0.17 & 19.99 $\pm$\ 0.28 & 18.52 $\pm$\ 0.08 & 17.34 $\pm$\ 0.02 & 16.33 $\pm$\ 0.02 & 16.27 $\pm$\ 0.02 & 16.3 $\pm$\ 0.02  & 16.32 $\pm$\ 0.02 \\
2152-53874-290 & 7557 $\pm$\ 40  & 4.14 $\pm$\ 0.23 & -1.02 $\pm$\ 0.09 & 19.07 $\pm$\ 0.16 & 17.47 $\pm$\ 0.05 & 16.17 $\pm$\ 0.01 & 15.11 $\pm$\ 0.02 & 15.11 $\pm$\ 0.01 & 15.17 $\pm$\ 0.01 & 15.26 $\pm$\ 0.02 \\
3406-54970-104 & 7191 $\pm$\ 81  & 4.32 $\pm$\ 0.02 & -1.98 $\pm$\ 0.02 & 21.27 $\pm$\ 0.32 & 19.92 $\pm$\ 0.11 & 18.77 $\pm$\ 0.02 & 17.87 $\pm$\ 0.02 & 17.79 $\pm$\ 0.02 & 17.77 $\pm$\ 0.03 & 17.83 $\pm$\ 0.02 \\
3406-54970-518 & 7335 $\pm$\ 50  & 4.25 $\pm$\ 0.18 & -1.79 $\pm$\ 0.04 & 21.91 $\pm$\ 0.44 & 19.89 $\pm$\ 0.11 & 18.8 $\pm$\ 0.02  & 17.9 $\pm$\ 0.02  & 17.8 $\pm$\ 0.01  & 17.82 $\pm$\ 0.01 & 17.86 $\pm$\ 0.03 \\
3387-54951-106 & 7979 $\pm$\ 116 & 4.21 $\pm$\ 0.38 & -1.84 $\pm$\ 0.21 & 20.53 $\pm$\ 0.4  & 20.07 $\pm$\ 0.2  & 18.75 $\pm$\ 0.02 & 17.73 $\pm$\ 0.02 & 17.73 $\pm$\ 0.02 & 17.87 $\pm$\ 0.02 & 17.99 $\pm$\ 0.02 \\
3384-54948-315 & 7403 $\pm$\ 67  & 3.89 $\pm$\ 0.33 & -1.92 $\pm$\ 0.07 & 21.29 $\pm$\ 0.28 & 19.45 $\pm$\ 0.07 & 18.29 $\pm$\ 0.02 & 17.3 $\pm$\ 0.02  & 17.24 $\pm$\ 0.02 & 17.25 $\pm$\ 0.02 & 17.32 $\pm$\ 0.02 \\
1348-53084-179 & 7652 $\pm$\ 15  & 4.19 $\pm$\ 0.21 & -2.24 $\pm$\ 0.0  & 19.6 $\pm$\ 0.18  & 18.53 $\pm$\ 0.06 & 17.59 $\pm$\ 0.02 & 16.59 $\pm$\ 0.01 & 16.59 $\pm$\ 0.02 & 16.67 $\pm$\ 0.02 & 16.69 $\pm$\ 0.02 \\
2124-53770-535 & 7005 $\pm$\ 44  & 4.08 $\pm$\ 0.04 & -1.45 $\pm$\ 0.05 & 21.25 $\pm$\ 0.45 & 18.57 $\pm$\ 0.09 & 17.13 $\pm$\ 0.02 & 16.25 $\pm$\ 0.01 & 16.12 $\pm$\ 0.01 & 16.12 $\pm$\ 0.01 & 16.15 $\pm$\ 0.01 \\
2336-53712-115 & 7469 $\pm$\ 22  & 4.07 $\pm$\ 0.23 & -0.6 $\pm$\ 0.07  & 21.01 $\pm$\ 0.4  & 19.96 $\pm$\ 0.19 & 18.5 $\pm$\ 0.02  & 17.46 $\pm$\ 0.02 & 17.4 $\pm$\ 0.01  & 17.45 $\pm$\ 0.01 & 17.52 $\pm$\ 0.02 \\
424-51893-60   & 7176 $\pm$\ 23  & 3.93 $\pm$\ 0.11 & -0.61 $\pm$\ 0.09 & 21.31 $\pm$\ 0.46 & 18.84 $\pm$\ 0.09 & 17.09 $\pm$\ 0.02 & 16.04 $\pm$\ 0.02 & 15.94 $\pm$\ 0.02 & 15.95 $\pm$\ 0.01 & 16.0 $\pm$\ 0.02  \\
2445-54573-173 & 7317 $\pm$\ 87  & 4.25 $\pm$\ 0.34 & -1.37 $\pm$\ 0.46 & 21.78 $\pm$\ 0.34 & 19.85 $\pm$\ 0.1  & 18.72 $\pm$\ 0.02 & 17.66 $\pm$\ 0.02 & 17.57 $\pm$\ 0.02 & 17.65 $\pm$\ 0.02 & 17.69 $\pm$\ 0.02 \\
1282-52759-79  & 7541 $\pm$\ 52  & 3.83 $\pm$\ 0.35 & -1.8 $\pm$\ 0.13  & 21.1 $\pm$\ 0.37  & 19.53 $\pm$\ 0.11 & 18.32 $\pm$\ 0.02 & 17.22 $\pm$\ 0.03 & 17.23 $\pm$\ 0.01 & 17.29 $\pm$\ 0.01 & 17.36 $\pm$\ 0.02 \\
3377-54950-189 & 7349 $\pm$\ 31  & 4.13 $\pm$\ 0.35 & -1.38 $\pm$\ 0.09 & 19.95 $\pm$\ 0.21 & 18.83 $\pm$\ 0.07 & 17.75 $\pm$\ 0.02 & 16.84 $\pm$\ 0.01 & 16.78 $\pm$\ 0.01 & 16.81 $\pm$\ 0.02 & 16.84 $\pm$\ 0.02 \\
2899-54568-252 & 7164 $\pm$\ 29  & 4.01 $\pm$\ 0.12 & -0.22 $\pm$\ 0.05 & 20.04 $\pm$\ 0.23 & 17.51 $\pm$\ 0.04 & 15.63 $\pm$\ 0.03 & 14.57 $\pm$\ 0.02 & 14.46 $\pm$\ 0.02 & 14.47 $\pm$\ 0.01 & 14.56 $\pm$\ 0.02 \\
1456-53115-620 & 7072 $\pm$\ 59  & 3.9 $\pm$\ 0.27  & -0.84 $\pm$\ 0.04 & 21.53 $\pm$\ 0.48 & 18.97 $\pm$\ 0.05 & 17.32 $\pm$\ 0.02 & 16.22 $\pm$\ 0.02 & 16.14 $\pm$\ 0.02 & 16.18 $\pm$\ 0.02 & 16.24 $\pm$\ 0.02 \\
2661-54505-400 & 7402 $\pm$\ 59  & 4.13 $\pm$\ 0.12 & -0.7 $\pm$\ 0.08  & 20.64 $\pm$\ 0.27 & 18.93 $\pm$\ 0.05 & 17.65 $\pm$\ 0.02 & 16.61 $\pm$\ 0.02 & 16.59 $\pm$\ 0.01 & 16.62 $\pm$\ 0.01 & 16.71 $\pm$\ 0.01 \\
335-52000-452  & 7758 $\pm$\ 159 & 4.1 $\pm$\ 0.16  & -2.03 $\pm$\ 0.06 & 19.58 $\pm$\ 0.19 & 19.11 $\pm$\ 0.1  & 18.49 $\pm$\ 0.02 & 17.54 $\pm$\ 0.02 & 17.61 $\pm$\ 0.01 & 17.66 $\pm$\ 0.01 & 17.7 $\pm$\ 0.02  \\
2647-54495-235 & 7096 $\pm$\ 51  & 3.9 $\pm$\ 0.12  & -1.47 $\pm$\ 0.02 & 21.61 $\pm$\ 0.42 & 19.04 $\pm$\ 0.09 & 17.62 $\pm$\ 0.02 & 16.66 $\pm$\ 0.02 & 16.56 $\pm$\ 0.02 & 16.55 $\pm$\ 0.02 & 16.59 $\pm$\ 0.02 \\
3253-54941-210 & 7296 $\pm$\ 41  & 3.96 $\pm$\ 0.27 & -1.74 $\pm$\ 0.04 & 21.47 $\pm$\ 0.47 & 19.29 $\pm$\ 0.07 & 17.99 $\pm$\ 0.02 & 17.06 $\pm$\ 0.02 & 17.0 $\pm$\ 0.02  & 17.03 $\pm$\ 0.02 & 17.08 $\pm$\ 0.02 \\
334-51993-267  & 7270 $\pm$\ 74  & 3.94 $\pm$\ 0.28 & -1.38 $\pm$\ 0.01 & 21.78 $\pm$\ 0.41 & 20.32 $\pm$\ 0.13 & 18.76 $\pm$\ 0.02 & 17.7 $\pm$\ 0.02  & 17.63 $\pm$\ 0.02 & 17.66 $\pm$\ 0.01 & 17.71 $\pm$\ 0.02 \\
3214-54866-426 & 7159 $\pm$\ 49  & 3.85 $\pm$\ 0.3  & -1.68 $\pm$\ 0.04 & 21.69 $\pm$\ 0.48 & 19.99 $\pm$\ 0.16 & 18.55 $\pm$\ 0.04 & 17.6 $\pm$\ 0.02  & 17.57 $\pm$\ 0.02 & 17.55 $\pm$\ 0.02 & 17.61 $\pm$\ 0.02 \\
2892-54552-128 & 7256 $\pm$\ 80  & 3.84 $\pm$\ 0.37 & -1.99 $\pm$\ 0.13 & 22.22 $\pm$\ 0.3  & 19.94 $\pm$\ 0.08 & 18.67 $\pm$\ 0.02 & 17.82 $\pm$\ 0.02 & 17.71 $\pm$\ 0.01 & 17.71 $\pm$\ 0.01 & 17.77 $\pm$\ 0.02 \\
3285-54948-131 & 7509 $\pm$\ 116 & 3.94 $\pm$\ 0.2  & -1.66 $\pm$\ 0.02 & 18.48 $\pm$\ 0.1  & 17.79 $\pm$\ 0.05 & 16.93 $\pm$\ 0.03 & 15.98 $\pm$\ 0.03 & 16.03 $\pm$\ 0.03 & 16.06 $\pm$\ 0.02 & 16.14 $\pm$\ 0.03 \\
2512-53877-326 & 7061 $\pm$\ 48  & 3.86 $\pm$\ 0.16 & -2.04 $\pm$\ 0.09 & 21.44 $\pm$\ 0.38 & 18.85 $\pm$\ 0.04 & 17.48 $\pm$\ 0.02 & 16.5 $\pm$\ 0.01  & 16.41 $\pm$\ 0.02 & 16.44 $\pm$\ 0.01 & 16.46 $\pm$\ 0.02 \\
3170-54859-111 & 7520 $\pm$\ 98  & 4.23 $\pm$\ 0.31 & -2.39 $\pm$\ 0.05 & 20.12 $\pm$\ 0.14 & 18.93 $\pm$\ 0.04 & 17.87 $\pm$\ 0.02 & 16.86 $\pm$\ 0.01 & 16.82 $\pm$\ 0.01 & 16.84 $\pm$\ 0.02 & 16.89 $\pm$\ 0.02 \\
878-52353-617  & 7240 $\pm$\ 51  & 4.14 $\pm$\ 0.31 & -1.81 $\pm$\ 0.16 & 20.61 $\pm$\ 0.3  & 18.79 $\pm$\ 0.07 & 17.41 $\pm$\ 0.02 & 16.44 $\pm$\ 0.02 & 16.33 $\pm$\ 0.02 & 16.36 $\pm$\ 0.03 & 16.38 $\pm$\ 0.02 \\
3193-54830-593 & 7838 $\pm$\ 78  & 4.19 $\pm$\ 0.22 & -1.14 $\pm$\ 0.02 & 20.44 $\pm$\ 0.22 & 19.71 $\pm$\ 0.11 & 18.66 $\pm$\ 0.03 & 17.55 $\pm$\ 0.02 & 17.62 $\pm$\ 0.01 & 17.66 $\pm$\ 0.02 & 17.75 $\pm$\ 0.03 \\
267-51608-422  & 7336 $\pm$\ 59  & 3.98 $\pm$\ 0.32 & -1.55 $\pm$\ 0.06 & 21.55 $\pm$\ 0.36 & 19.96 $\pm$\ 0.1  & 18.5 $\pm$\ 0.02  & 17.5 $\pm$\ 0.02  & 17.44 $\pm$\ 0.02 & 17.49 $\pm$\ 0.02 & 17.56 $\pm$\ 0.03 \\
1739-53050-128 & 7533 $\pm$\ 73  & 3.98 $\pm$\ 0.16 & -2.23 $\pm$\ 0.05 & 20.89 $\pm$\ 0.37 & 19.31 $\pm$\ 0.08 & 18.01 $\pm$\ 0.02 & 16.98 $\pm$\ 0.02 & 16.81 $\pm$\ 0.01 & 16.87 $\pm$\ 0.02 & 17.13 $\pm$\ 0.02 \\
1934-53357-4   & 7408 $\pm$\ 39  & 4.03 $\pm$\ 0.09 & -1.43 $\pm$\ 0.15 & 21.64 $\pm$\ 0.28 & 20.35 $\pm$\ 0.12 & 19.02 $\pm$\ 0.03 & 17.97 $\pm$\ 0.03 & 17.95 $\pm$\ 0.02 & 17.98 $\pm$\ 0.02 & 18.02 $\pm$\ 0.04 \\
1590-52974-388 & 7126 $\pm$\ 23  & 3.95 $\pm$\ 0.05 & -0.24 $\pm$\ 0.02 & 21.86 $\pm$\ 0.46 & 19.31 $\pm$\ 0.1  & 17.71 $\pm$\ 0.02 & 16.58 $\pm$\ 0.01 & 16.49 $\pm$\ 0.02 & 16.51 $\pm$\ 0.02 & 16.57 $\pm$\ 0.02 \\
3195-54832-552 & 7059 $\pm$\ 34  & 4.07 $\pm$\ 0.14 & -0.42 $\pm$\ 0.03 & 20.81 $\pm$\ 0.39 & 18.9 $\pm$\ 0.1   & 17.32 $\pm$\ 0.02 & 16.3 $\pm$\ 0.02  & 16.22 $\pm$\ 0.02 & 16.2 $\pm$\ 0.01  & 16.26 $\pm$\ 0.01 \\
3230-54860-409 & 7081 $\pm$\ 54  & 4.34 $\pm$\ 0.14 & -0.42 $\pm$\ 0.07 & 20.77 $\pm$\ 0.37 & 18.49 $\pm$\ 0.08 & 16.74 $\pm$\ 0.02 & 15.65 $\pm$\ 0.02 & 15.55 $\pm$\ 0.02 & 15.58 $\pm$\ 0.02 & 15.63 $\pm$\ 0.02 \\
2273-53709-207 & 7353 $\pm$\ 128 & 4.05 $\pm$\ 0.32 & -0.48 $\pm$\ 0.13 & 20.97 $\pm$\ 0.4  & 19.41 $\pm$\ 0.12 & 18.1 $\pm$\ 0.02  & 16.86 $\pm$\ 0.02 & 16.72 $\pm$\ 0.01 & 16.67 $\pm$\ 0.01 & 16.62 $\pm$\ 0.01 \\
3230-54860-353 & 7541 $\pm$\ 69  & 3.89 $\pm$\ 0.27 & -1.52 $\pm$\ 0.05 & 21.47 $\pm$\ 0.49 & 19.89 $\pm$\ 0.17 & 18.76 $\pm$\ 0.02 & 17.75 $\pm$\ 0.01 & 17.71 $\pm$\ 0.02 & 17.77 $\pm$\ 0.02 & 17.84 $\pm$\ 0.02 \\
3226-54857-452 & 8004 $\pm$\ 116 & 4.15 $\pm$\ 0.17 & -0.52 $\pm$\ 0.17 & 18.15 $\pm$\ 0.11 & 17.89 $\pm$\ 0.02 & 17.58 $\pm$\ 0.02 & 16.66 $\pm$\ 0.01 & 16.74 $\pm$\ 0.01 & 16.86 $\pm$\ 0.01 & 16.99 $\pm$\ 0.02 \\
758-52253-5    & 7930 $\pm$\ 117 & 3.99 $\pm$\ 0.18 & -1.34 $\pm$\ 0.04 & 18.47 $\pm$\ 0.12 & 18.45 $\pm$\ 0.04 & 18.45 $\pm$\ 0.03 & 17.75 $\pm$\ 0.02 & 17.79 $\pm$\ 0.02 & 17.92 $\pm$\ 0.01 & 18.05 $\pm$\ 0.03 \\
2420-54086-395 & 7780 $\pm$\ 101 & 4.07 $\pm$\ 0.38 & -1.47 $\pm$\ 0.05 & 21.25 $\pm$\ 0.35 & 20.17 $\pm$\ 0.12 & 18.82 $\pm$\ 0.02 & 17.84 $\pm$\ 0.01 & 17.8 $\pm$\ 0.01  & 17.87 $\pm$\ 0.01 & 17.94 $\pm$\ 0.02 \\
2541-54481-384 & 7282 $\pm$\ 26  & 3.87 $\pm$\ 0.0  & -0.24 $\pm$\ 0.01 & 20.22 $\pm$\ 0.21 & 18.0 $\pm$\ 0.04  & 16.32 $\pm$\ 0.02 & 15.26 $\pm$\ 0.01 & 15.16 $\pm$\ 0.03 & 15.17 $\pm$\ 0.02 & 15.25 $\pm$\ 0.02 \\
2055-53729-37  & 7297 $\pm$\ 44  & 3.99 $\pm$\ 0.25 & -0.56 $\pm$\ 0.05 & 19.97 $\pm$\ 0.17 & 17.86 $\pm$\ 0.04 & 16.37 $\pm$\ 0.01 & 15.32 $\pm$\ 0.02 & 15.26 $\pm$\ 0.0  & 15.27 $\pm$\ 0.01 & 15.35 $\pm$\ 0.01 \\
3166-54830-409 & 7203 $\pm$\ 14  & 4.18 $\pm$\ 0.12 & -0.45 $\pm$\ 0.03 & 20.72 $\pm$\ 0.34 & 18.28 $\pm$\ 0.07 & 16.59 $\pm$\ 0.02 & 15.55 $\pm$\ 0.01 & 15.46 $\pm$\ 0.01 & 15.49 $\pm$\ 0.02 & 15.61 $\pm$\ 0.02 \\
3161-54779-445 & 7351 $\pm$\ 62  & 4.04 $\pm$\ 0.23 & -0.81 $\pm$\ 0.1  & 21.42 $\pm$\ 0.48 & 19.45 $\pm$\ 0.13 & 18.2 $\pm$\ 0.02  & 17.13 $\pm$\ 0.02 & 17.11 $\pm$\ 0.01 & 17.22 $\pm$\ 0.02 & 17.3 $\pm$\ 0.02  \\
2941-54507-638 & 7097 $\pm$\ 25  & 4.09 $\pm$\ 0.11 & -0.46 $\pm$\ 0.02 & 20.89 $\pm$\ 0.27 & 18.37 $\pm$\ 0.04 & 16.64 $\pm$\ 0.02 & 15.65 $\pm$\ 0.01 & 15.51 $\pm$\ 0.01 & 15.52 $\pm$\ 0.01 & 15.57 $\pm$\ 0.02 \\
2337-53740-222 & 7100 $\pm$\ 26  & 3.89 $\pm$\ 0.17 & -0.24 $\pm$\ 0.04 & 19.51 $\pm$\ 0.17 & 17.98 $\pm$\ 0.06 & 16.27 $\pm$\ 0.01 & 15.17 $\pm$\ 0.01 & 15.04 $\pm$\ 0.01 & 15.04 $\pm$\ 0.01 & 15.1 $\pm$\ 0.01  \\ \hline
\end{tabular}%
}
\end{table}
\label{lastpage}

\end{document}